\documentclass[
    ,final            
  ]
  {aipproc}

\layoutstyle{6x9}
\newcommand{\be}{\begin{eqnarray}}
\newcommand{\ee}{\end{eqnarray}}
\newcommand{\kT}{{\bf k}_\perp}
\newcommand{\bT}{{\bf b}_\perp}

\begin{document}

\title{Hadron Tomography}


\author{Matthias Burkardt}{
  address={Department of Physics, New Mexico State University,
Las Cruces, NM 88003, U.S.A.}
}



\begin{abstract}
Generalized parton distributions describe the distribution
of partons in the transverse plane. For transversely polarized
quarks and/or nucleons, these impact parameter dependent parton
distributions are not axially symmetric. These transverse
distortions can be related to spin-orbit correlations as well as to (intuitively) to transverse single-spin asymmetries,
allowing novel insights into quark orbital angular momentum
from measurements of the Sivers and Boer-Mulders functions.
\end{abstract}

\maketitle


\section{Introduction}
\begin{figure}
\unitlength1.cm
\begin{picture}(10,3.7)(1,10.3)
\includegraphics{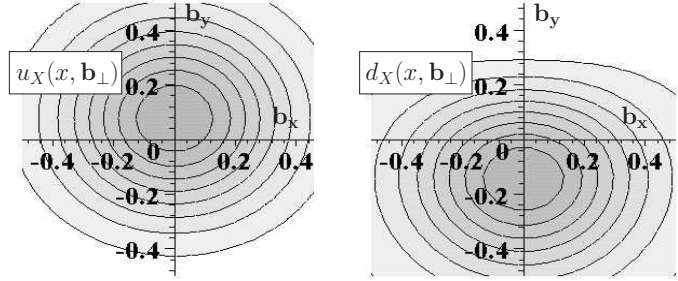}
\end{picture}
\caption{Expected impact parameter dependent PDF for
$u$ and $d$ quarks ($x_{Bj}=0.3$ is fixed) for a proton
polarized in the $+\hat{x}$ direction in the model from
Ref. \cite{GPDe}.
}
\label{fig:distort}
\end{figure}

Generalized parton distributions (GPDs) \cite{JR} for $\xi =0$
provide information
about the distribution of partons in impact parameter space
The distribution of  unpolarized quarks in unpolarized nucleons  
is given
by the Fourier transform of the GPD $H(x,0,-\Delta_\perp^2)$ \cite{GPD,diehl}
\be
q(x,\bT)= \int \frac{{\rm d}^2\Delta_\perp^2}{(2\pi)^2}
H(x,0,-\Delta_\perp^2) e^{-i\Delta_\perp\cdot\bT},
\ee
where $H$ is the GPD which appears in a decomposition
of the Dirac form factor w.r.t. the momentum
$x$ of the active quark $F_1(t)=\int {\rm d}x H(x,\xi,t)$.
The reference point for ${\bf b}_\perp$ is the $\perp$ center
of longitudinal momentum ${\bf R}_\perp = \sum_{i\in q,g}
x_i {\bf r}_{\perp,i}$ \cite{soper}.
A similar relation exists for longitudinally
polarized quarks $\Delta q(x,\bT)$ with ${H}(x,0,-\Delta_\perp^2)
\longrightarrow \widetilde{H}(x,0,-\Delta_\perp^2)$. For a $\perp$
polarized target the distribution of partons in impact parameter 
space
is no longer axially symmetric and the deformation is described by 
the Fourier transform of the GPD
$E(x,0,-\Delta_\perp^2)$.
For example, for a nucleon polarized in the $+\hat{x}$ direction,
one finds \cite{GPDe}
\be
q(x,\bT)= \int \frac{{\rm d}^2\Delta_\perp^2}{(2\pi)^2}
H(x,0,-\Delta_\perp^2) e^{-i\Delta_\perp\cdot\bT}
- \frac{1}{2M}\frac{\partial }{ \partial b_y}
\int \frac{{\rm d}^2\Delta_\perp^2}{(2\pi)^2}
E(x,0,-\Delta_\perp^2) e^{-i\Delta_\perp\cdot\bT}
\label{E}
\ee
(see Fig. \ref{fig:distort}).
Little is known about the GPD $E$, beyond that it provides a $x$
decomposition
of the Pauli form factor as $F_2(t)=\int {\rm d}x E(x,\xi,t)$.
Eq. (\ref{E}) yields
a $\perp$ flavor dipole moment
\be
d_y = \int {\rm d}^2\bT q(x,\bT) \bT = \frac{\kappa_{q/p}}{2M}, 
\ee
where $\kappa_{q/p}$ is the contribution from quark Flavor $q$ to the
anomalous magnetic moment of the nucleon (with the charge factored
out). Neglecting the small contribution from strange quarks, the
anomalous magnetic moments of the proton and neutron are sufficient
to perform a flavor decomposition of the anomalous magnetic
moment, yielding
$\kappa_{u/p}=1.67$ and $\kappa_{d/p}=-2.03$, i.e.
significant $\perp$ flavor dipole moments $|d_q| \approx 0.2 fm$.

The physical origin for this deformations is the orbital motion
of the quarks. When viewed in the Breit frame, quarks with orbital
angular momentum in the transverse direction move towards the
virtual photon on one side of the nucleon and away from it on the
other side, i.e. they are shifted towards larger momentum fractions
on the side of the nucleon where they move towards the virtual 
photon. As PDFs are rapidly falling functions of $x$, the resulting
increase in $x$ for quarks on one side results in an
enhancement when viewed at fixed $x$ and the PDFs
appear shifted towards that side. Due to the opposite signs for
$\kappa_{u/p}$ and $\kappa_{d/p}$, the shifts are into opposite
directions for $u$ and $d$ quarks, indicating opposite signs for 
their orbital angular momenta.

\section{Sivers Function}
\begin{figure}
\unitlength1.cm
\begin{picture}(10,2.3)(3.,19.2)
\includegraphics{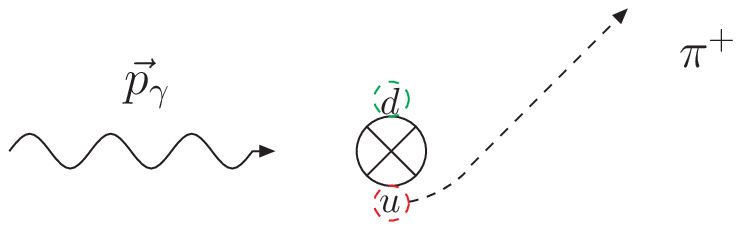}
\end{picture}
\caption{The transverse distortion of the parton cloud for a proton
that is polarized into the plane, in combination with attractive
FSI, gives rise to a Sivers effect for $u$ ($d$) quarks with a
$\perp$ momentum that is on the average up (down).}
\label{fig:deflect}
\end{figure}

The significant distortion of parton distributions in impact 
parameter space is expected to have observable consequences in
other experiments. For example, in semi-inclusive DIS, when the 
virtual photon strikes a $u$ quark in a $\perp$ polarized proton,
the $u$ quark distribution is enhanced on the left side of the target
(for a proton with spin pointing up when viewed from the virtual 
photon perspective). Although in general the final state 
interaction (FSI) is very complicated, we expect it to be on average attractive thus translating a position space
distortion to the left into a momentum space asymmetry to the right
and vice versa (Fig. \ref{fig:deflect}).
Such a $\perp$ spin asymmetry is usually
parameterized in terms of the Sivers function 
$f_{1T}^{\perp q}(x,{\bf k}_\perp^2)$ entering the unintegrated
parton density in a $\perp$ polarized target 
\cite{sivers,mauro,trento}
\be
f_{q/p^\uparrow}(x,\kT)=f_1^q(x,{\bf k}_\perp^2)
-f_{1T}^{\perp q}(x,{\bf k}_\perp^2)
\frac{(\hat{\bf P}\times \kT) \cdot{\bf S}}{M}
\ee
Although this picture is very intuitive, a few words
of caution are in order. First of all the above mechanism is strictly
true only in mean field models for the FSI as well as in simple
spectator models \cite{spectator}. 
Furthermore, even in such mean field models
there is no one-to-one correspondence between quark distributions
in impact parameter space and unintegrated parton densities
(e.g. Sivers function). Although both are connected by a Wigner
distribution \cite{wigner}, they are not Fourier transforms of each other.
Nevertheless, since the primordial momentum distribution of the quarks
(without FSI) must be symmetric we find a qualitative connection
between the primordial position space asymmetry and the
momentum space asymmetry (with FSI). 
Another issue concerns the $x$-dependence of the Sivers function.
The $x$-dependence of the position space asymmetry is described
by the GPD $E(x,0,-{\Delta}_\perp^2)$. Therefore, within the above
mechanism, the $x$ dependence of the Sivers function should be
related to the $x$ dependence of $E(x,0,-{\Delta}_\perp^2)$, for 
example, in a mean field model for the FSI, 
the second moment of the
Sivers function is obtained as
$\int {\rm d}^2{\bf k}_\perp^2 f_{1T}^{\perp q}(x,{\bf k}_\perp^2)
{\bf k}_\perp^2 \propto 
\int {\rm d}^2{\bf b}_\perp {\cal E}(x,{\bf b}_\perp)
 {\bf \nabla}\cdot {\bf I}({\bf b}_\perp)$, where 
${\bf I}({\bf b}_\perp)$ describes the FSI \cite{lensing}.
However, the $x$ dependence of $E$ is not known yet and we only
know the Pauli form factor $F_2=\int {\rm d}x E$. Nevertheless, 
if one makes
the additional assumption that $E$ does not fluctuate as a function 
of $x$ then the contribution from each quark flavor $q$ to the
anomalous magnetic moment $\kappa$ determines the sign of 
$E^q(x,0,0)$
and hence of the Sivers function. Making these assumptions,
as well as the very plausible assumption that the FSI is on average
attractive (${\bf \nabla}\cdot {\bf I}({\bf b}_\perp)\leq 0$), 
one finds that $f_{1T}^{\perp u}<0$, while 
$f_{1T}^{\perp d}>0$. Both signs have been confirmed by the HERMES
experiment \cite{hermes}.

For the flavor analysis of the Sivers function, there exists a useful
sum-rule, which states that the average ${\bf k}_\perp$
summed over all 
$x$, and all quarks and gluons vanishes \cite{sumrule}
\be
\sum_{i\in q,g} \int {\rm dx} \int {\rm d}^2{\bf k}_\perp 
f_{1T}^{\perp i}(x,{\bf k}_\perp^2)
{\bf k}_\perp^2 = 0.
\label{sumrule}
\ee
However, Eq. (\ref{sumrule}) is {\it not} a trivial consequence 
of $\perp$ momentum conservation in the SIDIS
process, as the different $f_{1T}^{\perp i}$ are extracted from
different experiments. To illustrate this point, 
we consider a gedanken 
experiment using the scalar diquark model as a target. Imagine first
a probe that only couples to the fermion $f$ in the model. The
$2^{nd}$ moment of $f_{1T}^{\perp f}$ would be evaluated by measuring
the $\perp$ momentum of the fermions produced in a DIS experiment
with a probe that couples only to the fermion. Now consider a second
probe which couples only to the scalar in this model. This time only
the $\perp$ momenta of the scalars are measured in the final state.
The sum rule (\ref{sumrule}) states that the $2^{nd}$ moment of the
fermion  $\perp$ momentum distribution in the first experiment plus
the $2^{nd}$ moment of the scalar $\perp$ momentum distribution in 
the second experiment should add up to zero. In contradistinction,
mere momentum conservation yields that the $\perp$ momenta
of {\it all} constituents in the final state (both active and 
spectator)
in one {\it single} experiment should add up to zero. 
Eq. (\ref{sumrule}) is a different statement than that.
Present experiments \cite{hermes,compass} seem to indicate that
the sum rule is already close to being satisfied by summing over 
$u$ and $d$ quarks only, suggesting possibly a small Sivers effect
for the glue \cite{gardner}.  

\section{Chirally odd GPDs and the Boer-Mulders Function}
\subsection{Transversity Distribution in Impact Parameter
Space}
\begin{figure}
\unitlength0.90cm
\begin{picture}(15,5.9)(-10.5,0.7)
\includegraphics{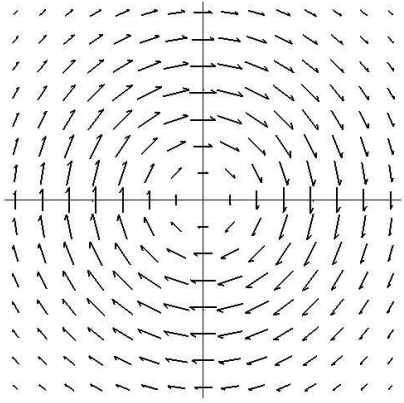}
\end{picture}
\caption{Distribution of transversity in impact parameter space
 (schematical).}
\label{wirbel}
\end{figure}

Even in an unpolarized target, the distribution of quarks with a 
certain transverse spin may not be axially symmetric either as
the transverse quark polarization singles out a direction as well.
The details of this distortion are described by the chirally odd
GPD $\bar{E}_T(x,0,-{ \Delta}_\perp^2)$ \cite{DH}. 
Physically, what happens
here is that although all orientations for the quark orbital motion
are equally likely, there may be a certain preferred correlation
between the quark orbital motion and the quark spin (e.g. due to
spin-orbit effects). Therefore, by looking at distributions of 
quarks with a certain transversity one does at the same time
single out (at least preferentially) quarks with a certain 
orientation of the orbital motion (Fig. \ref{wirbel}). 
Here the sign of the correlation depends on
the sign of the spin-orbit correlation, which is one of the
interesting pieces of information that we are trying to find 
out.    
Once again, there will be an 
``enhancement'' of PDFs on the side of the nucleon where the
quarks move
towards the virtual photon. This enhancement manifests itself
in a distortion sideways relative to the (transverse) quark spin, 
which is described by the GPD $\bar{E}_T$. The distortion is 
$\perp$ to the quark spin and its sign is determined by the 
sign of $\bar{E}_T$. When quark with a certain $\perp$ polarization
are knocked out in a DIS experiment, the momentum distribution of
these quarks in the final state thus exhibits a left-right 
asymmetry analogous to the asymmetry described by the Sivers
function, except that the $\perp$ spin of the nucleon is replaced
by the $\perp$ spin of the struck quark \cite{BM}
\be
f_{q^\uparrow/p}(x,\kT)=\frac{1}{2}\left[f_1^q(x,{\bf k}_\perp^2)
-{h_{1}^{\perp q}(x,{\bf k}_\perp^2)}
\frac{(\hat{\bf P}\times \kT)
\cdot{{\bf S}_q}}{M}\right].\label{eq:BM}
\ee
Using the Collins effect \cite{collins}, i.e. 
the $\perp$ asymmetry in the distribution of hadrons in a jet
originating from a transversely polarized quark, one can
`tag' the spin of the active quark: relative to the lepton 
scattering plane the pion distribution in the final state exhibits 
a $\cos (2\phi)$ asymmetry proportional to the Collins as well
as the `Boer-Mulders function' $h_{1}^{\perp q}(x,{\bf k}_\perp^2)$,
which can thus be extracted. 
 
\subsection{Are All Boer-Mulders Functions Alike?}
While the signs of the chirally even GPD $E^q$ are known from the 
flavor decomposition of the anomalous magnetic moment, no such 
information is available for the chirally odd GPD $\bar{E}^q$. 
Therefore, the only
information on $\bar{E}^q$ comes from models and lattice studies.
We considered a bag model for both the nucleon as well as the
pion, as well as the NJL model for the pion \cite{brian}. 
In all of these
calculations we find that $\kappa_T=\int {\rm d}x \bar{E}^q>0$ 
for both $q=u,d$. Moreover, a constituent quark model 
\cite{pasquini},
using a Melosh rotation to generate the spin-orbit structure
of the light-cone wave functions, also yields $\kappa_T^u>0$ and
$\kappa_T^d>0$ for the nucleon. All of these models also
yirlded numerical values for $\kappa_T$ that were larger than for
the (chirally even) $\kappa$
Of course, all of these are just model results ond one may question
their validity. Fortunately, moments of $\int dx \bar{E}^q$
have been calculated in lattice gauge theory and were also found 
to be positive and numerically larger than $\kappa$.

Similar to the chirally even GPD $E$, $\bar{E}_T$ also requires
the interference between wave function components that differ
by one unit of quark orbital angular momentum. In the bag model,
the ground state wave function only contains nonzero orbital
angular momentum in its lower component, which is related to the
upper component through the free Dirac equation. In the constituent
quark model the situation is similar. At least in those models the
sign of $\bar{E}_T$ can thus be understood from the relative phase
between upper and lower components in $s$-wave solutions to the
free Dirac equation \cite{brian}. 
Amazingly, lattice QCD yields the same signs \cite{hagler,schafer}.
The same sign for $\kappa_T^u$ and $\kappa_T^d$ can also
be understood on the basis of large $N_C$, as for $N_C\rightarrow
\infty$ one finds $\bar{E}_T^u=\bar{E}_T^d$.

The fact that $\kappa_T$ is larger than $\kappa$ can also be 
understood in these models (bag model and constituent quark model), 
which have in common that the nucleon wave function is constructed
from a product of independent quark wave functions, which one would
also do in a mean field approximation appropriate for 
$N_C\rightarrow\infty$. What these models/approximations have in 
common is that the nucleon wave function is constructed from
independent quark wave functions. However, regardless whether
they which although these independent quark wave functions have $J_z = +\frac{1}{2}$ or $J_z = -\frac{1}{2}$, both yield the same
correlation between the quark spin and the quark orbital motion 
as both the spin and the orbital motion
are reversed in the $J_z= -\frac{1}{2}$ wave function. When one 
constructs the nucleon state with total angular momentum an spin
equal to $\frac{1}{2}$, for both $u$ and $d$ quarks 
this involves both quark wave functions with $J_z=+\frac{1}{2}$
as well as $J_z= -\frac{1}{2}$. In the computation of
the GPD $E$, which is sensitive to the correlation between
quark orbital angular momentum and the {\it nucleon} spin, the
involvement of quark wave functions with both $J_z=\pm \frac{1}{2}$,
results in a certain cancellation between different terms in
the familiar $SU(6)$ wave functions used to construct the nucleon
state. However, in the case of $\bar{E}_T$, no such cancellation
occurs, as each quark wave function has the same correlation between
its orbital angular momentum and its spin, and all $SU(6)$
components contribute coherently to $\bar{E}_T$.
As a result, there is no cancellation and these models all 
yield $\kappa_T>\kappa$, and in fact
all models where the nucleon state is constructed from products
of identical (up to $J_z$) quark wave functions, e.g. mean field
models, should share this property. Again, lattice calculations 
give results that are consistent with these models. 

\section{Summary}
Parton distributions in impact parameter space are $\perp$
deformed when either the nucleon is $\perp$ polarized or when 
one considers the distribution or $\perp$ polarized quarks.
These deformations provide a simple mechanism for $\perp$
SSAs, based on the assumption that the FSI is 
on average attractive. The predicted signs for the Sivers functions 
for $u$ and $d$ quarks were confirmed in a recent experiment 
\cite{hermes}. Various models as well as lattice QCD calculations
predict that the lowest moment of $\bar{E}_T$ is positive
for both $u$ and $d$ quarks and numerically larger than the 
lowest moment of its chirally even counterpart $E$.
As a result, the Boer-Mulders 
function for $u$ as well as $d$ quarks is expected to be larger than 
the Sivers function for the same flavor and 
with the same sign as the Sivers function for $u$ quarks, i.e.
negative. 
%

\begin{theacknowledgments}
  This work was in part supported by the DOE (DE-FG03-95ER40965).
\end{theacknowledgments}




\begin{thebibliography}{9}



\bibitem{JR} D. M\"uller et al., Fortschr. Phys. {\bf 42}, 101
(1994); X.~Ji, Phys. Rev. Lett. {\bf 78}, 610 (1998); 
A.V.~Radyushkin, Phys. Rev. {\bf D 56}, 5524.

\bibitem{GPD} M. Burkardt, Phys. Rev. {\bf D 62}, 071503 (2000);
Erratum-ibid. {\bf D 66}, 119903 (2002).

\bibitem{diehl} M. Diehl, Eur. Phys. J. {\bf C 25}, 223 (2002);
J.P.~Ralston and B.~Pire, Phys. Rev. D {\bf 66}, 111501
(2002).

\bibitem{soper} D. Soper, Phys. Rev. {\bf D 15}, 1141 (1977).

\bibitem{GPDe} M. Burkardt, Int. J. Mod. Mhys. {\bf A 18}, 173 (2003).
\bibitem{lensing} M. Burkardt, Nucl. Phys. {\bf A 735}, 185 (2004);
Phys. Rev. {\bf D 69}, 057501 (2004). 

\bibitem{sivers} D.W. Sivers, Phys. Rev. {\bf D 43}, 261 (1991).

\bibitem{mauro} M. Anselmino, M. Boglione, and F. Murgia,
Phys.\ Rev.\ D {\bf 60}, 054027 (1999); V. Barone, A. Drago, and
P.G. Ratcliffe, Phys. Rept. {\bf 359}, 1 (2002);
M. Anselmino, U.D'Alesio, and F. Murgia, Phys.\ Rev.\ D {\bf 67},
074010 (2003).

\bibitem{trento} A. Bacchetta et al., Phys. Rev. {\bf D 70}, 117504
(2004).

\bibitem{spectator} S.J. Brodsky, D.S. Hwang, and I. Schmidt,
Phys. Lett. {\bf B 530}, 99 (2002); M. Burkardt and D.S. Hwang,
Phys. Rev. {\bf D 69}, 074032 (2004); Z. Lu and I. Schmidt, 
hep-ph/0611158.


\bibitem{wigner} A.V. Belitsky, X. Ji, and F. Yuan, Phys. Rev. {\bf D 69}, 074014 (2004).

\bibitem{hermes} HERMES collaboration, Phys. Rev. Lett.
{\bf 94}, 012002 (2005).

\bibitem{sumrule} M. Burkardt, Phys. Rev. {\bf D 69}, 091501
(2004).

\bibitem{compass} COMPASS Collaboration (E.S. Ageev et al.),
hep-ex/0610068.

\bibitem{gardner} S.J. Brodsky and S. Gardner, hep-ph/0608219.

\bibitem{DH} M. Diehl and P. H\"agler, Eur. Phys. J. {\bf C 44},
87 (2005).

\bibitem{BM} D. Boer and P.J. Mulders, Phys. Rev. {\bf D 57}, 5780 (1998).

\bibitem{collins} J.C.~Collins, Phys. Lett. {\bf B 536}, 43 (2002)
CHECK!


\bibitem{hagler} QCDSF-UKQCD Collaboration, M. G\"ockeler et al.,
PoS LAT2005, 055 (2006).

\bibitem{pasquini} B. Pasquini, M. Pincetti, and S. Boffi,
Phys. Rev. {\bf D 72}, 094029 (2005).

\bibitem{schafer} A. Sch\"afer, talk given at ECT*, Trento, June 2006.


\bibitem{brian} M. Burkardt and B. Hannafious,  to be published.

\end{thebibliography}
\end{document}